# Localized in-situ polymerization on graphene surfaces for stabilized graphene dispersions

Sriya Das[*], Ahmed S. Wajid[*], John L. Shelburne[*], Yen-Chi Liao[*], Micah J. Green[*%]

[*]*Department of Chemical Engineering, Texas Tech University, Lubbock, Texas 79409, USA*

[%]*Corresponding author: micah.green@ttu.edu*

We demonstrate a novel *in situ* polymerization technique to develop localized polymer coatings on the surface of dispersed pristine graphene sheets. Graphene sheets show great promise as strong, conductive fillers in polymer nanocomposites; however, difficulties in dispersion quality and interfacial strength between filler and matrix have been a persistent problem for graphene-based nanocomposites, particularly for pristine graphene.  With this in mind, a physisorbed polymer layer is used to stabilize graphene sheets in solution. To form this protective layer, an organic microenvironment is formed around dispersed graphene sheets in surfactant solutions, and a nylon 6, 10 or nylon 6, 6 coating is created via interfacial polymerization. This technique lies at the intersection of emulsion and admicellar polymerization; a similar technique was originally developed to protect luminescent properties of carbon nanotubes in solution. These coated graphene dispersions are aggregation-resistant and may be reversibly redispersed in water even after freeze-drying. The coated graphene holds promise for a number of applications, including multifunctional graphene-polymer nanocomposites.

Single-layer graphite, known as graphene, has attracted considerable scientific interest in recent years because graphene's unique mechanical, electrical, and thermal properties may enable a range of advanced materials and devices including nanocomposites [1-5], thin conductive films [6,7]. Graphene is a two-dimensional structure of $sp^2$-hybridized carbon atoms arranged in a honeycomb lattice [8]. Graphene was originally isolated through micromechanical cleavage of graphite; this discovery was recently honored with the 2010 Nobel Prize in Physics [9,10]. However, this method is limited due to the lack of scalability. Recent advancements in the production of graphene involve chemical vapor deposition [11], epitaxial growth, and liquid phase exfoliation from graphite. This latter method is best suited to the scalable production of multifunctional advanced materials (such as composites) based on graphene.

However, dispersion of graphene in common solvents is challenging because exfoliation from graphite is hindered by the strong, attractive van der Waals forces holding the sheets together. Even after the initial process of exfoliation and dispersion, those same attractive forces cause graphene to reaggregate. We briefly review the common techniques used to address this issue.

The most common technique for exfoliation and dispersion of graphene is the oxidation of graphite to form graphite oxide [12-14]. The graphite oxide is hydrophilic and is easily exfoliated in water and other solvents as single sheets, termed graphene oxide (GO). The presence of the carboxyl and epoxide groups on the basal plane of GO reduce the interlayer forces and render them soluble in water. (Note that recent work has suggested that GO is not actually soluble in water and other solvents; instead, oxidative debris acts as a stabilizer [15].) GO may be reduced using hydrazine in the presence of stabilizers such as surfactants or polymers to yield chemically converted graphene (CCG) [12]. The stabilizers wrap around the graphene and sterically prevent reaggregation [13,14]. Additional functional groups may be bonded to GO or CCG to increase solubility in a range of solvents [16-19]. However, this approach suffers from certain drawbacks. The reduction of GO to CCG is incomplete such that some of the $sp^3$ characteristics of GO are still retained in the CCG [12]. This process only partially restores the unique properties of pristine graphene; in fact, the electrical conductivity of CCG is two orders of magnitude lower than pristine graphene [20].

Alternatively, liquid phase exfoliation of graphene as few-layer sheets may be obtained in certain organic solvents without chemical modification [6,21]. The use of organic solvents such as NMP yields defect-free monolayer graphene dispersions [22], but the concentrations of the dispersions are comparatively low (< 0.01 mg/ml) and require extensive sonication. Chlorosulfonic acid acts as an excellent solvent for graphene and yields graphene dispersions with concentrations as high as 2 mg/ml; unfortunately, such superacids are incompatible with most composite applications [23]. Intercalation compounds such as potassium intercalants are used to increase the distance in between the consecutive layers of graphite; these intercalation compounds aids in the exfoliation of graphene without functionalization or sonication [24,25].

Aqueous dispersions of pristine graphene may be prepared by sonicating graphite in the presence of stabilizers; such stabilizers include polymers, such as poly-vinyl pyrrolidone (PVP), or surfactants, such as sodium chlorate and sodium dodecyl benzene sulfonate (SDBS)[26] [27-33]. The surfactant technique yields high concentration dispersions of graphene (> 0.3 mg/ml) [32].

For composite applications, excellent dispersion is not enough; one must also have excellent interfacial adhesion for efficient stress transfer from the graphene to the polymer matrix. These two issues of poor dispersion and poor interfacial strength between filler and matrix is a persistent problem for composites with pristine nanomaterial fillers [34]. Composites of graphene nanoribbons (GNR)[35] and graphene platelet[36,37] with an epoxy matrix have shown a considerable increase in mechanical properties; however, the graphene platelets used in these composites are either GO or CCG since dispersion of pristine graphene is challenging. Furthermore, surfactant-free graphene is desirable since presence of surfactant may affect the transparency, thermal properties, and mechanical properties of the composite [38-41]. The ability to form polymer coatings on pristine graphene surfaces may address this issue of interfacial strength.

In the present work, we aim to address these needs; we utilize an *in situ* polymerization technique to encapsulate the graphene sheets with a polymer coating. Our work lies at the intersection of emulsion polymerization (where polymerization occurs at the organic-aqueous interface inside micelles in the bulk solution) and admicellar polymerization (where polymerization occurs at a surface coated by a surfactant bilayer). Admicellar polymerization refers to the process in which a thin film of polymer is deposited on a solid surface. This polymerization process is observed to go through three major steps: (1) the aggregation of surfactants at solid-liquid interfaces to form bilayer (admicelles) through adsorption from an aqueous solution (2) introduction of the hydrophobic monomer which travels to the inner-core of the surfactant (monomer adsolubilization) and (3) initiation of the polymerization by mechanisms similar to those occurring in emulsion techniques. In case of the admicellar polymerization, the surfactant concentration is maintained to be always below the critical micelle concentration (cmc) to avoid polymerization in the bulk solution. This technique was used to coat various substrates with polymers like polystyrene (PS) and poly methyl methacrylate (PMMA) [42-46]. This coating technique was also applied to study the method of polymer coating formation on graphite and mica surfaces [47-51]. Chen *et al.* had previously demonstrated that the micelle surrounding single-walled carbon nanotubes (SWNTs) could be swelled by various organic solvents [52]. They then developed a technique to coat surfactant-stabilized SWNTs with nylon 6,10 by interfacial polymerization within the micelle to preserve SWNT optical properties in solution[53]. We hypothesized that such a technique could be retrofitted to graphene dispersions despite the substantial differences in surfactant structure between stabilized SWNTs and graphene. This work is a combination of both emulsion and admicellar polymerization, as the graphene is inside the micelle and a thin polymer coating is deposited on the graphene surface. Whereas Chen *et al.* focused on SWNT optical properties, our ultimate application is graphene-based composites. The polymer-coating should blend well with the polymer matrix and facilitate stress transfer and transport of properties in the composite; this issue of load transfer is one of the more pressing problems in the field of nanocomposites.

We first swell the surfactant micelle around the graphene with organic solvents, and polymerization performed at the interface between water and organic solvents creates a nylon 6,10 coating on the graphene surface. (This technique can be used to create nylon 6,6 coatings as well.) The coated graphene can be freeze-dried and redispersed effectively in water. The coating aids the redispersion of graphene in water without any visible aggregation by forming a polymer layer around the graphene and preventing van der Waals induced aggregation. We characterize our dispersions through a variety of techniques, including rheology, absorbance, AFM, TEM and SEM imaging. The polymer-coated graphene holds great promise as a means to increase the interfacial strength in polymer nanocomposites.

**Experimental Procedure:**

Stable dispersions of graphene in water are prepared with sodium dodecyl benzene sulfonate (SDBS) (MPBiochemicals, # 157889) as the surfactant. Expanded graphite (1.29 grams), (Asbury Carbons, CAS# 7782-42-5, Grade 3805) is added to a 15ml (2%w/v) solution of SDBS (The use of graphite flakes instead of expanded graphite yields similar results). The use of 2 % w/v ratio (20 mg/ml) of surfactant is adopted from the work of Green *et al.* , where sodium cholate is used to stabilize graphene in water.[30] The critical micelle concentration of SDBS is usually found to be 2.9 mM (1 mg/ml) at 298 K[54]. This solution is tip sonicated using a Misonix sonicator (XL 2000) at output wattage of 7W for 1 hour. The sample is further centrifuged (Centrific Centrifuge 225, Fischer Scientific) overnight (12 hours) at a speed of ~ 5000 rpm to remove larger aggregates, and the supernatant is collected.

A 0.5M sebacoyl chloride (Sigma Aldrich, 99%) solution in carbon tetrachloride (Acros Organics, 99%) is used to swell the micelle structure over the graphene. (In order to make nylon 6,6 rather than nylon 6,10, adipoyl chloride is used rather than sebacoyl chloride.) The surfactant/graphene dispersion is carefully added to the organic solution in a 1:1 ratio, shaken and allowed to stand for one hour for phase separation. The organic solvent forms an interface and swells the micelle interior. After phase separation, the swelled aqueous surfactant/graphene dispersion is separated from the suspension. Hexamethylene diamine (Sigma Aldrich, 98%) is melted at 50°C and 2 μl is added to the swelled aqueous dispersion drop wise using a micropipette. Hexamethylene diamine and sebacoyl chloride react at the water/organic interface to form nylon on the surface of the graphene.

To measure the concentration of the dispersions, the samples are filtered through Teflon filters (Millipore, 0.2μm), dried overnight at 40°C and the change of mass of the filter paper was measured carefully. Based on prior TGA analysis[31], we estimate that ~64% of the residue on the filter paper is graphitic. (The same filter papers are also used to measure Raman spectra on a Renishaw Raman microscope using a 633nm He-Ne laser.) To correlate concentration with absorbance, UV-vis spectroscopy is performed on a Shimadzu UV-vis spectrophotometer 2550 at wavelengths of 200nm to 800nm. To eliminate the background effect, i.e. the effects of surfactant spectrum, the absorbance is measured against the surfactant solution.

As a check for dispersion stability at low pH, the pH of the surfactant-stabilized graphene dispersion is lowered by adding 1.2M HCl solution to it and centrifuging the sample at ~5000 rpm for one hour; this is directly compared against the low pH nylon-coated graphene dispersion.

As an additional check for stability against van der Waals aggregation, both the surfactant-stabilized graphene and the nylon-coated graphene dispersions may be freeze dried (Vitris Benchtop Freeze Dryer) overnight to yield dry samples. The freeze-dried samples are redispersed in water without any sonication for rheological and other analysis. Rheological experiments are done using a double Couette fixture (C-DG26.7/T200/SS) on a shear rheometer (Anton Paar, USA). To prevent the evaporation of the samples, a solvent trap is used. The shear-viscosity behavior of aqueous samples was measured before freeze drying and after freeze drying and redispersion.

The structure of the graphite flakes and the nylon-coated graphene is characterized by FT-IR spectroscopy (Nexus 470).For FTIR analysis, ethyl acetate (EA) (5ml) is added to the nylon-coated samples and further freeze-dried (EA removes the surfactant from the system). After phase separation the nylon/graphene portion is carefully removed from the EA portion.

Tapping Mode AFM analysis is done on a Veeco Multimode AFM (IIIa) with NSC14 cantilevers (MikroMasch). AFM samples are prepared by spin coating ~20µl of the dispersion onto a freshly cleaved mica surface at 3000 rpm for 20 seconds. The spin coated mica is further dried on a hot plate at 50 °C for 1 minute.

TEM samples are prepared by drop coating the sample on holey carbon grids and air-dried for 1 minute. A voltage of 100 keV is used to image the specimens on a Hitachi H8100.

SEM samples are prepared by mounting the samples on double-faced carbon tape and sputter coating with gold at 10 mA current for 1 minute. A voltage of 2 keV is used to image the specimens on a Hitachi S4300 SE/N.

**Results and Discussions:**

The schematic representation (Fig. 1a) shows the basic mechanism of polymerization and coating around the graphene. First, the graphite flakes are sonicated in an SDBS solution. (The ability of SDBS to stabilize graphene dispersions is well established [31].) The sonicated sample is then centrifuged to yield a dark solution with no precipitate (Fig. 1b). Similar to the SWNT-nylon technique, the surfactant environment is swelled with organic solvent [53]. Carbon tetrachloride (with dispersed sebacoyl chloride) is added to the surfactant-stabilized graphene dispersion which penetrates into the gap between the surfactant and graphene. This creates a water-organic interface between the graphene surface and the surfactant which is ideal for interfacial polymerization since the monomers for nylon formation are selectively soluble in the water (hexamethylene diamine) and the organic phase (sebacoyl chloride). The hexamethylene diamine is added to the aqueous phase and reacts with sebacoyl chloride at the interface. This creates a thin nylon coating on the graphene pictured in Fig 1c. The stepwise mechanism of the emulsion polymerization is demonstrated in Fig. 1d. In the case of admicellar polymerization, the organic monomer is typically added to the aqueous phase. In our work, we use an organic solvent to aid the transportation of the organic monomer to the hydrophobic core of the surfactant micelles.

We utilize a range of characterization techniques to establish that (1) we begin with a high-concentration surfactant-stabilized few-layer graphene dispersion, (2) the nylon polymerization is successful and the dispersion remains stable, and (3) the nylon protects the graphene from van der Waals aggregation in scenarios (low pH, freeze drying) where simple surfactant stabilization would not. This final point is especially critical because the creation of aggregation-resistant graphene opens up a wide range of novel applications for graphene, particularly in the field of composite melt-mixing.

The concentrations of the centrifuged phase of the graphene dispersions are determined by vacuum filtration. In the case of the SDBS/graphene dispersions, the concentration is found to be 0.2 mg/ml (comparable to 0.09-0.3 mg/mL for sodium cholate [55,56] and 0.05 mg/mL for SDBS [31]). These concentrations are quantitatively correlated with the absorbance spectra. According to Lambert-Beer's law, the absorption coefficient of any substance varies linearly with the concentration. The absorbance spectra of the dispersions of surfactant-stabilized graphene, solvent-swelled surfactant stabilized graphene and surfactant-stabilized nylon-coated-graphene are provided in the supplementary information (Fig S1). Fig. S2 in the supplementary information shows the optical absorbance as a function of different concentrations of the graphene dispersions. We determined the extinction coefficient ($α$) at a wavelength of 660nm using the linear relationship between the absorbance and calculated concentration for a particular dispersion ($A = αlC$; where $l$ is the cell length). Absorbance at 660nm wavelength was used by Lotya *et al.* to calculate the extinction coefficients for the same system.[31,57] In case of our SDBS/graphene dispersion, $α$ is found to be 1660 mL mg$^{-1}$ m$^{-1}$ which is in reasonable agreement with the values of extinction coefficient values (1390 mL mg$^{-1}$ m$^{-1}$) obtained by Lotya *et al.*[31].

The degree of exfoliation is measured by Raman spectroscopy. The parent expanded graphite (Fig. 2a) shows a very sharp G peak and comparatively smaller 2D peak at 2700 cm$^{-1}$. In the spectra of surfactant-stabilized (Fig. 2b) and polymer-coated graphene (Fig. 2c) the intensity of the D-peak increases. Prior studies have shown that as the graphene flake size decreases, the number of graphene edges exposed per flake increases [22,23,30,33,58 23,58]. These edges have sp$^3$ characteristics which contribute to the increases in D-peak intensity. The 2D peak position for graphene is 3-5cm$^{-1}$ shifted relative to the parent graphite[58]. The SDBS/graphene and the polymer-coated graphene show a G-peak shift (~3cm$^{-1}$), which indicates exfoliation of graphene. The 2D peak ideally should be around four times as intense as the G-peak for monolayer graphene[58]. In Fig. 2c, the intensity of the 2D peak has a larger intensity compared to the G peak, which depict that the coated graphene is a few layers thick. Fig. 2d shows a comparison of all the three samples tested. These observations confirm the exfoliation of few-layer graphene.

FT-IR spectroscopy is performed on the freeze dried samples of surfactant-stripped, nylon-coated graphene and parent graphite flakes to investigate the chemical change caused by the polymerization [53]. Similar to the nylon-coated SWNTs, the nylon-coated graphene also exhibits characteristic amide-I, amide-II, N-H and C-H stretches in the FT-IR spectrum. Figure 3 shows spectral comparison where the transmittance of graphite is shifted by a constant, for clarity. The polymer-coated graphene sample shows distinct peaks of C-H stretching at 1220cm$^{-1}$ and 2920cm$^{-1}$, the amide-II peaks at 1450cm$^{-1}$, and a broad peak of the N-H stretching at 3410cm$^{-1}$. A comparison between the spectra of graphite flakes and polymer-coated graphene confirms that nylon formation occurs in the bulk solution similar to Chen *et al*.

Fig. 4 shows SEM images of the freeze-dried samples. The surfactant-stabilized sample and nylon-coated graphene sample exhibit a strong morphological difference. The coated sample (Fig. 4b) can be identified by the difference of the appearance in the image. The magnified views of the freeze dried samples are shown in Fig. 4c & 4d. More SEM images of the vacuum filtered films and ethyl acetate treated vacuum filtered films of the surfactant stabilized sample and nylon-coated graphene are provided in the supplementary information (Fig. S3).

To access the number of layers of the graphene in the dispersions, tapping mode AFM analysis is performed on the samples. An AFM image of surfactant-stabilized graphene is shown in Fig. 5a. Prior work has shown that the graphene in the surfactant-stabilized aqueous graphene dispersions is indeed single-to-few layers thick[30,31]. In our images, we also observe that the surfactant stabilizes single-to-few layer graphene. The height ranges from 1-2.5 nm. The surfactant is deposited on the graphene in an irregular fashion. The layer of surfactant is found to have a patchy appearance. The presence of the surfactant on single-to-few layer graphene accounts for the variation in thickness. Fig. 5b shows the graphene flakes after interfacial polymerization. The thickness variation in this case is 4-6 nm. The polymerization changes the morphology of the surface without causing aggregation. The thin coating of nylon on the graphene causes the increase in thickness. Prior work has shown that the polymer usually gets deposited on the substrate in the form of 'patches' or islands. AFM measurements on mica and graphite surfaces reveal that the polymer film formation is indeed discontinuous[47,51,59,60]. In case of the AFM images of our polymer-stabilized graphene, we observe an irregular deposition of the polymer on the graphene surface. The AFM images of irregular polymer deposition are similar to those of See *et al.* and Marquez *et al.*[50,60], who observed polymerization on graphite surfaces. For completeness, more results on the AFM performed on the nylon-coated surfactant-stabilized graphene are provided in the supplementary information (Fig. S4). The nylon-coated graphene can be easily redispersed after freeze drying, and to check the dispersion quality, we take AFM images after redispersion (Fig. 5c). There is no change in thickness of the graphene flakes after redispersion. This gives an evidence of an excellent redispersion of coated-graphene in water without aggregation. We revisit this issue below by making rheological measurements.

We also use TEM to confirm the presence of single-to-few layer stabilized graphene in our dispersions. A TEM image of the nylon-coated graphene is shown in Fig. 6a. The edge of a graphene sheet can reveal the the number of layers present. Fig. 6b shows the SDBS/nylon/graphene dispersion, and the 2-3 layered edges indicate that the graphene is 2-3 layers thick. This is similar to our AFM results on the SDBS/nylon/graphene. Additional TEM images of both SDBS/nylon/graphene and SDBS/graphene are available in supplementary information (Fig. S5).

As we investigate the differences in dispersion stability caused by the polymerization, we note that the interfacial polymerization reaction generates an acidic pH [53]. The measured pH of the nylon-coated surfactant-stabilized graphene is found to be in the range of 1.7~2.5. Prior work showed graphene-PAM dispersions were stable with no aggregation down to a pH of 4 [61]. If the pH of our SDBS/graphene dispersions is lowered to the same range as the nylon-coated graphene dispersions, the SDBS/graphene dispersion destabilizes very easily (Fig. 7a) while the nylon/graphene dispersion remains stable (Fig. 7b); this proves that the polymer actually stabilizes graphene at the low pH, similar to the nanotubes results of Chen *et al*. [53]. The coated graphene can thus be processed into graphene-based materials and devices at varying pH conditions.

Additionally, the surfactant stabilized sample (Fig. 7c-d) and the nylon-coated sample (Fig. 7e-f) can be freeze dried and redispersed in water without any sonication. Although both the redispersed samples look alike to the eye, we wish to quantify any differences in dispersion quality. Note that in the case of SWNTs, a well-resolved fluorescence spectrum with strong peaks proves that the SWNTs remain as individuals [53,62]; thus, the fluorescence spectra of SWNTs, combined with the absorbance and Raman spectra, give a measure of the dispersion quality. Such a technique is not feasible for graphene due to the absence of fluorescence; instead, we rely on rheological measurements to characterize the dispersion quality before freeze-drying and after redispersion (Fig. S6).

These results may have implications for the processing of graphene-based nanocomposites. In most prior studies, melt mixing of pristine graphene into a polymer matrix has met little success due to difficulties in exfoliation and stable dispersion. Even if the graphene is well-dispersed in the polymer matrix, there still tends to be poor interfacial strength between graphene and polymer matrix, i.e., the load transfer is poor from the matrix to the high-strength filler. For our system, if the nylon-coating treatment were applied to graphene, this coated graphene could easily be melt mixed into a bulk nylon matrix since the coating prevents van der Waals contact between graphene sheets; furthermore, the physisorbed nylon coating should enhance load transfer between the graphene and the surrounding nylon matrix. Preliminary data on such nylon composites loaded with nylon-coated graphene composites are included in supplementary information (Fig S7). We aim to explore these issues in future studies.

In summary, we have developed a simple and scalable method to exfoliate graphene in water. In contrast to GO-based methods, no oxidation and reduction are involved in our procedure. We successfully coat the graphene non-covalently with nylon 6,10 and nylon 6,6, and the stability of the graphene dispersion in water is enhanced.

**Acknowledgements:**

We acknowledge Colin Young and Professor Matteo Pasquali of Rice University for their help with the Raman measurements. We acknowledge Wei Zheng and Professor Sindee Simon of TTU for assisting in the FT-IR spectra measurements. We acknowledge Gina Paroline of Anton Paar for her unique insight and help with the rheological measurements. The SEM was performed at the TTU Imaging Center funded by (NSF MRI 04-511) supported by Dr Mark J. Grimson and Professor Lauren S. Gollahon. We thank Professor Brandon Weeks of TTU for his expertise and equipment used in the AFM experiments. We thank Dr. Huipeng Chen and Professor Ronald Hedden of TTU for their helpful insights on nylon composites. Funding was provided by National Science Foundation (NSF) under award CBET-1032330.

Figures:

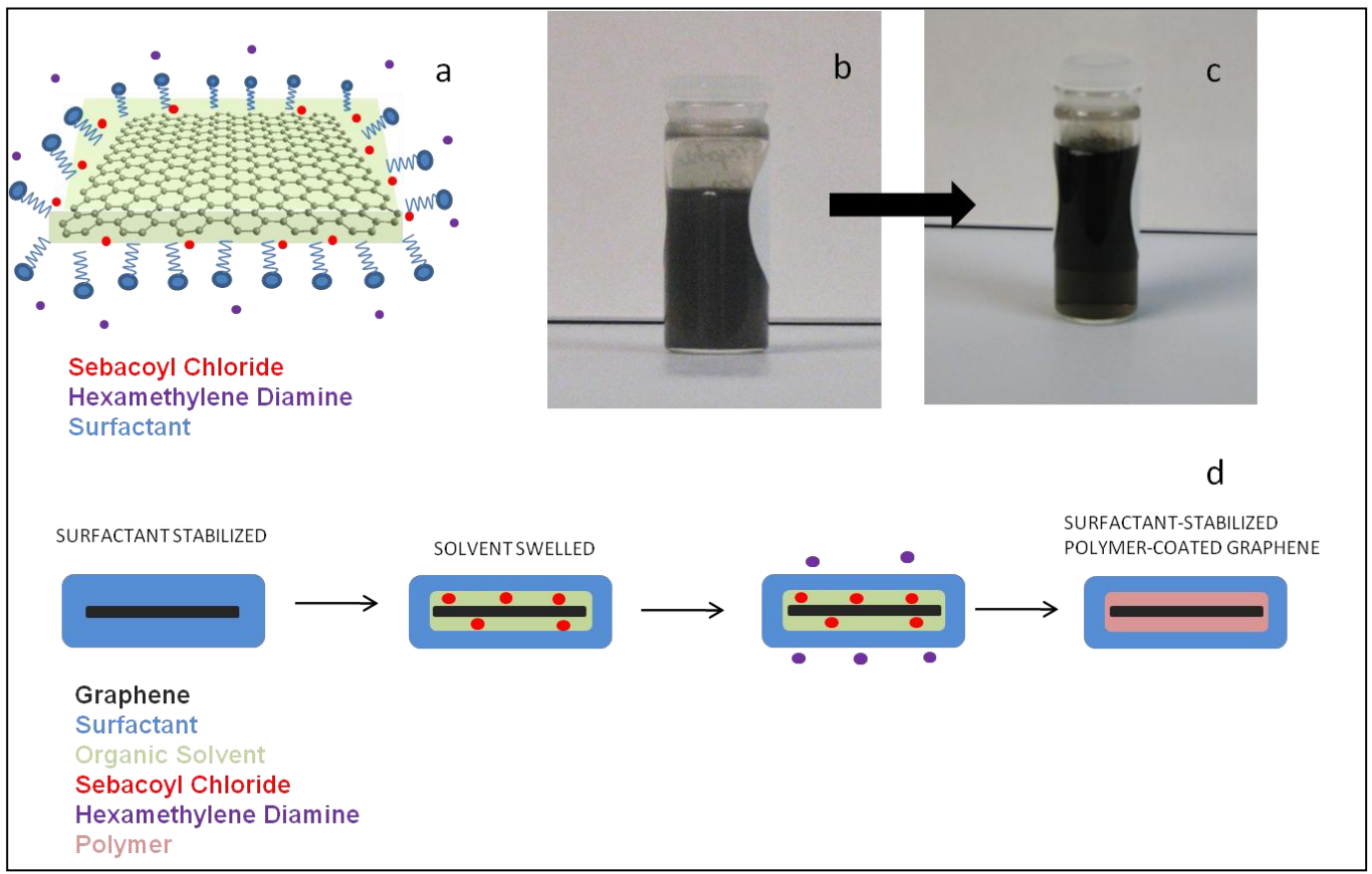

Fig 1: (a) Schematic representation of mechanism of wrapping of nylon around graphene. Photograph of (b) Aqueous dispersion of surfactant stabilized graphene. (c) Aqueous dispersion of nylon-coated surfactant stabilized graphene. (d) Schematic representation of the polymerization technique to stabilize graphene.

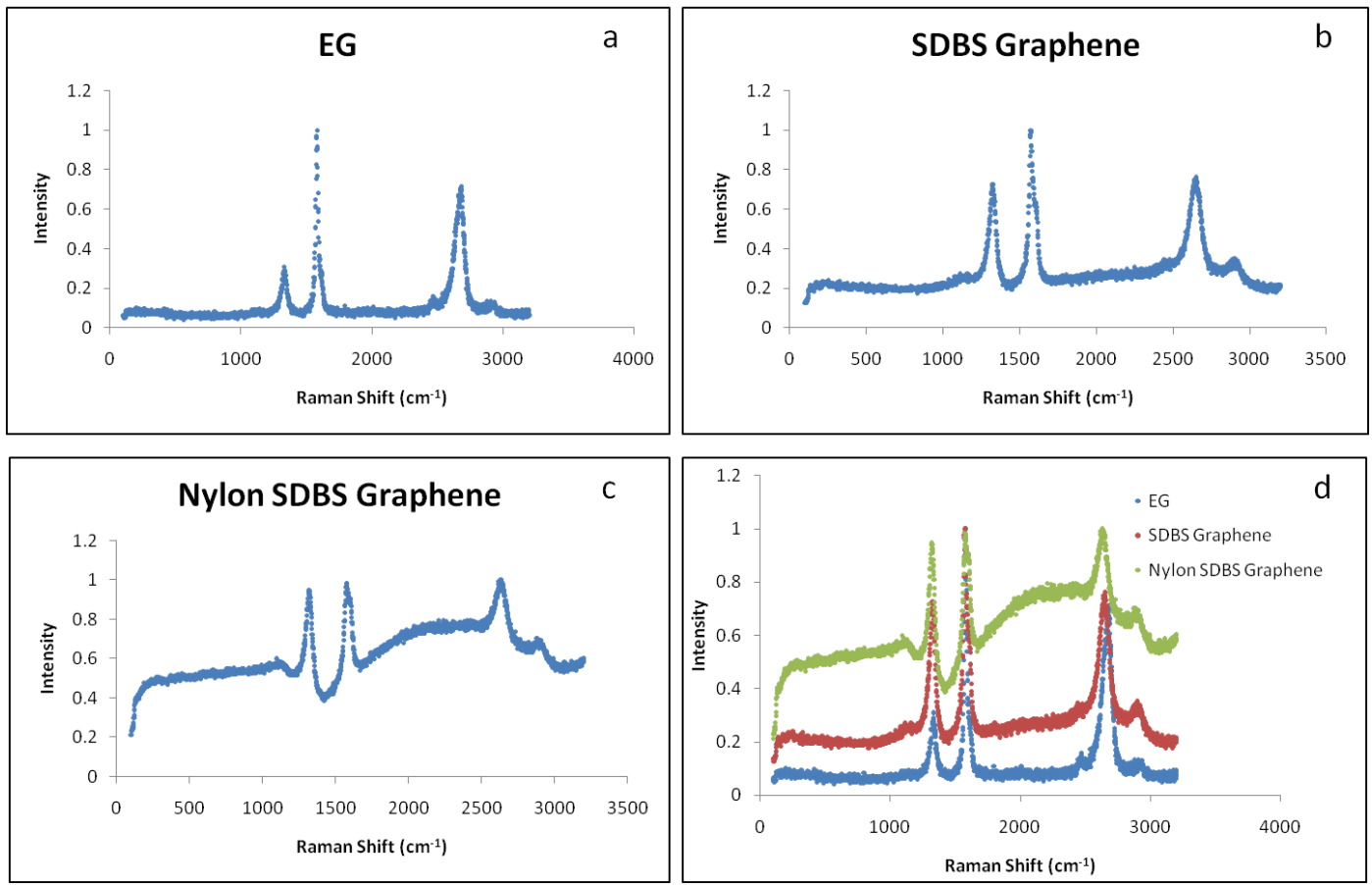

Fig 2: Raman spectra of (a) parent graphite (expanded graphite) (EG), (b) SDBS/graphene, (c) nylon/SDBS/graphene and (d) the comparison of all the spectra. In the spectra of surfactant-stabilized and polymer-coated graphene, we observe an increase in the intensity of the D-peak. As the graphene flake size decreases, the number of edges of the graphene exposed per flake increases. The sp$^3$ characteristic of the edges contributes to the intensity of the D-peak. The exfoliation of SDBS/graphene and the polymer-coated graphene is confirmed by a G-peak shift (~3cm$^{-1}$).

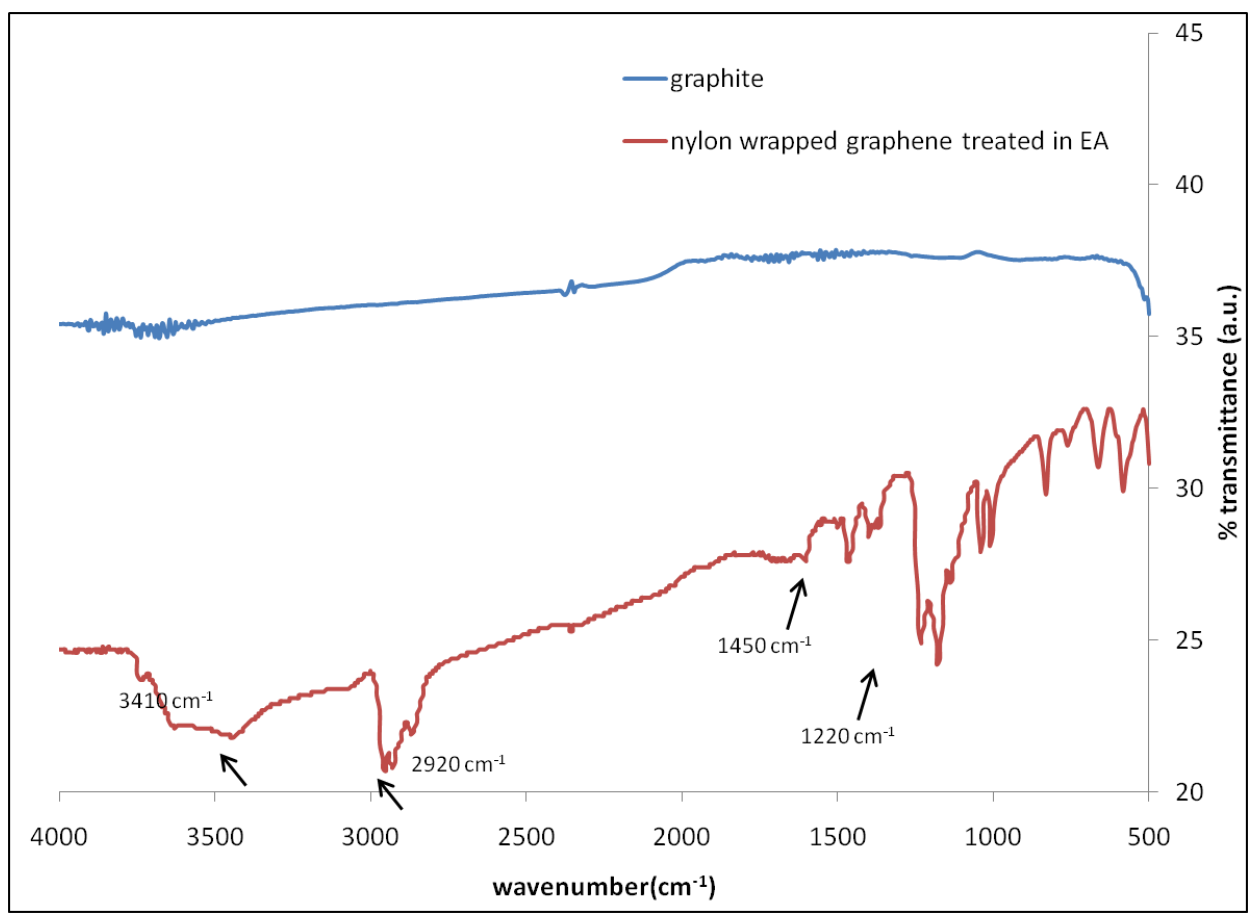

Fig 3: FTIR spectra of raw and nylon/graphene. The effect of the surfactant is removed by treating the coated graphene with ethyl acetate (EA). The distinct peaks of C-H stretching at 1220cm$^{-1}$ and 2920cm$^{-1}$, the amide-II peaks at 1450cm$^{-1}$, and a broad peak of the N-H stretching at 3410cm$^{-1}$ in case of the polymer-coated graphene confirms the nylon formation.

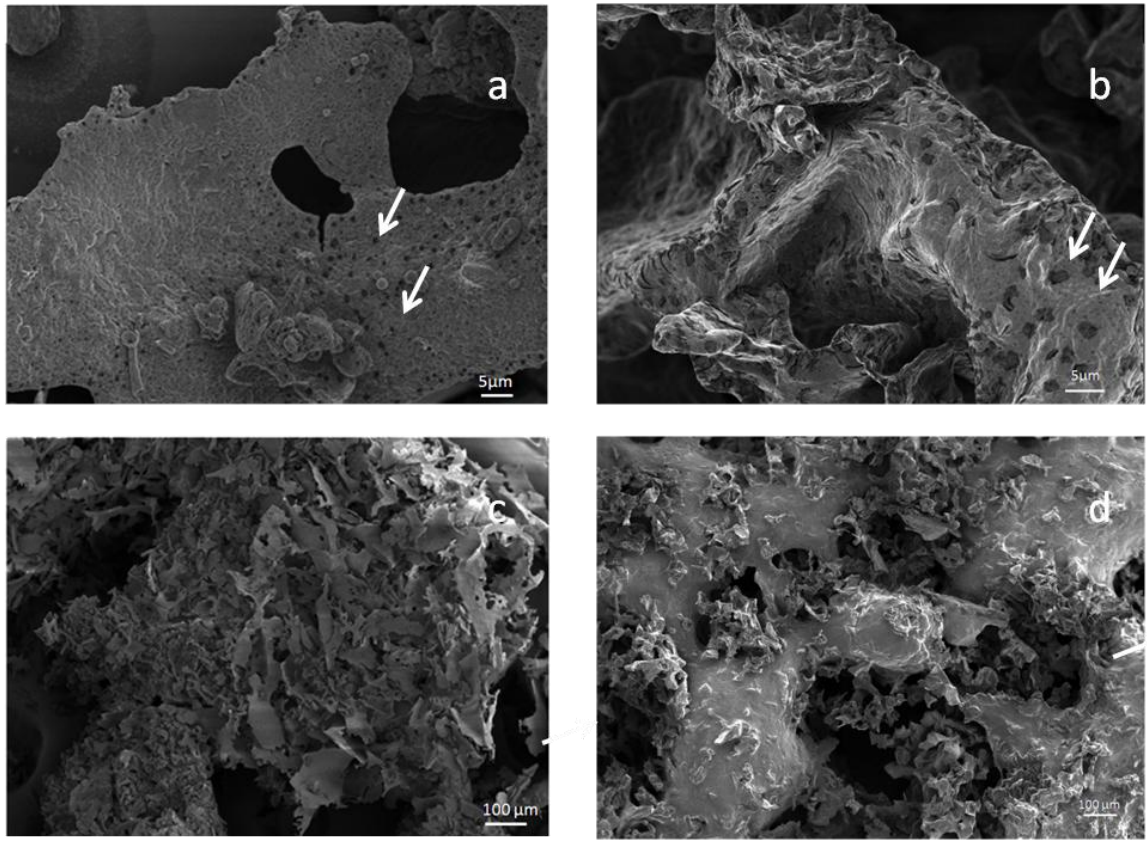

Fig 4: SEM images on freeze-dried samples of (a) SDBS/graphene (b) nylon/SDBS/graphene; magnified views of freeze-dried samples of (c) SDBS/graphene (d) nylon/SDBS/graphene. The surfactant-stabilized and nylon-coated surfactant-stabilized graphene samples show no visible aggregation under the EM studies.

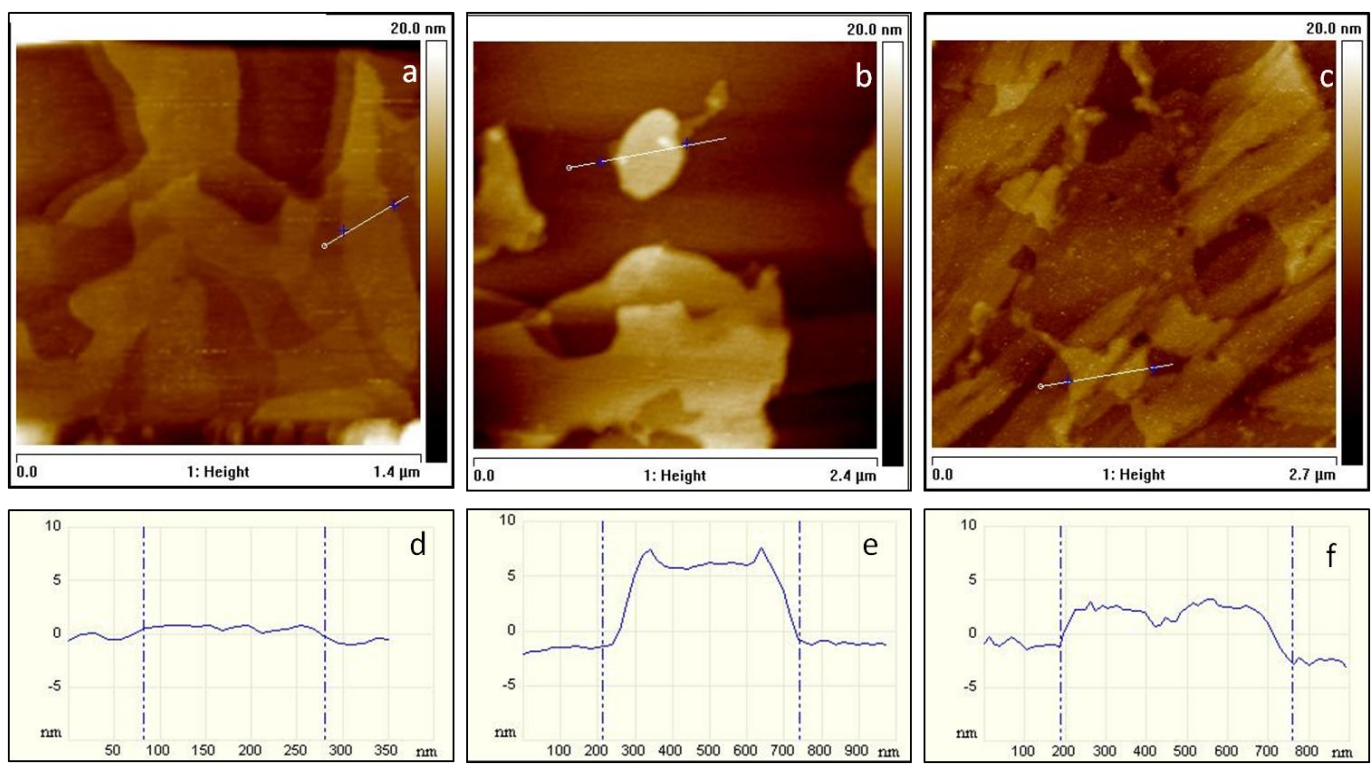

Fig 5: AFM of the (a) SDBS/graphene, (b) nylon/SDBS/graphene, (c) redispersed nylon/SDBS/graphene (d) height profile of the flake in SDBS/graphene, (e) height profile of the flake in nylon/SDBS/graphene, (f) height profile of the flake in redispersed nylon-coated SDBS-stabilized graphene. In case of the surfactant-stabilized graphene, the height ranges from 1-4 nm indicating that the graphene is indeed single-to-few layer thick. The nylon-coated graphene shows a height variation of 4-6nm. The thin coating of nylon on the surfactant-stabilized graphene surface increases the thickness without aggregation. AFM images of the nylon-coated graphene after freeze drying and redispersion shows no change in thickness of the graphene. This implies excellent redispersion of the coated graphene after freeze drying.

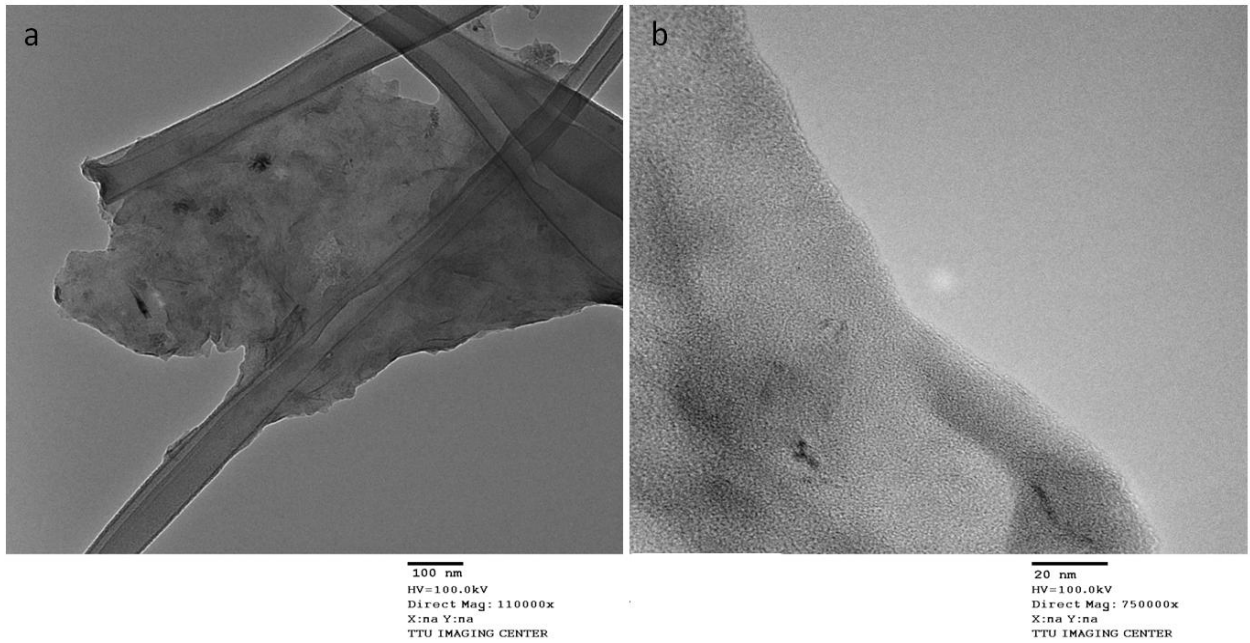

Fig 6: TEM images of (a) nylon/SDBS/graphene; magnified views of (b) nylon/SDBS/graphene. The magnified view of the edges indicates that the graphene is 2-3 layers thick.

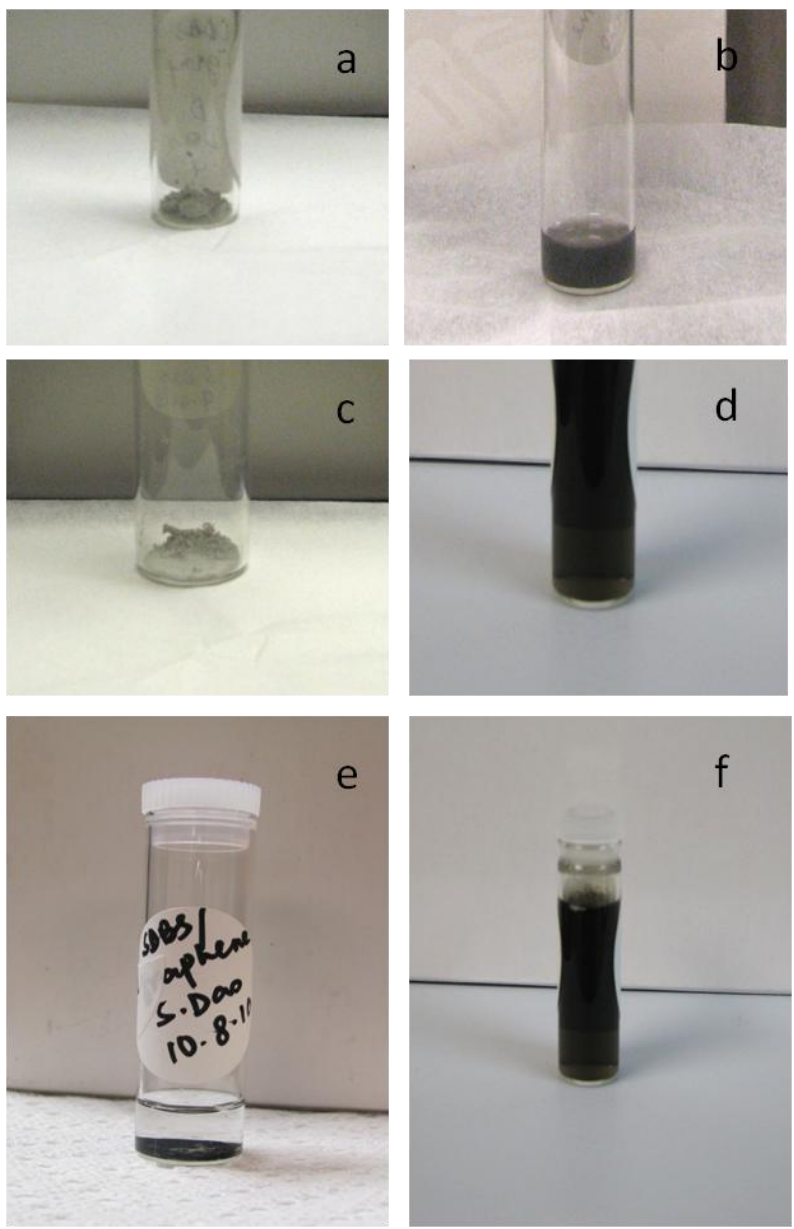

Fig 7: Digital camera images of SDBS/graphene (a) freeze dried; (b) redispersed in water; nylon/SDBS/graphene (c) freeze dried; (d) redispersed in water; (e) SDBS/Graphene dispersion at a pH of 2.5; (f) Nylon/SDBS/Graphene at a pH of 2.5. The interfacial polymerization reaction generates an acidic pH. The pH of the nylon-coated surfactant-stabilized graphene lies in the range of 1.7~2.5The pH of the SDBS/graphene dispersion is lowered to the same range as the nylon-coated graphene dispersion by adding HCl for the sake of comparison. The SDBS/graphene dispersion destabilizes (Fig 7e) easily proving that the polymer stabilizes graphene even at the low pH (Fig 7f).

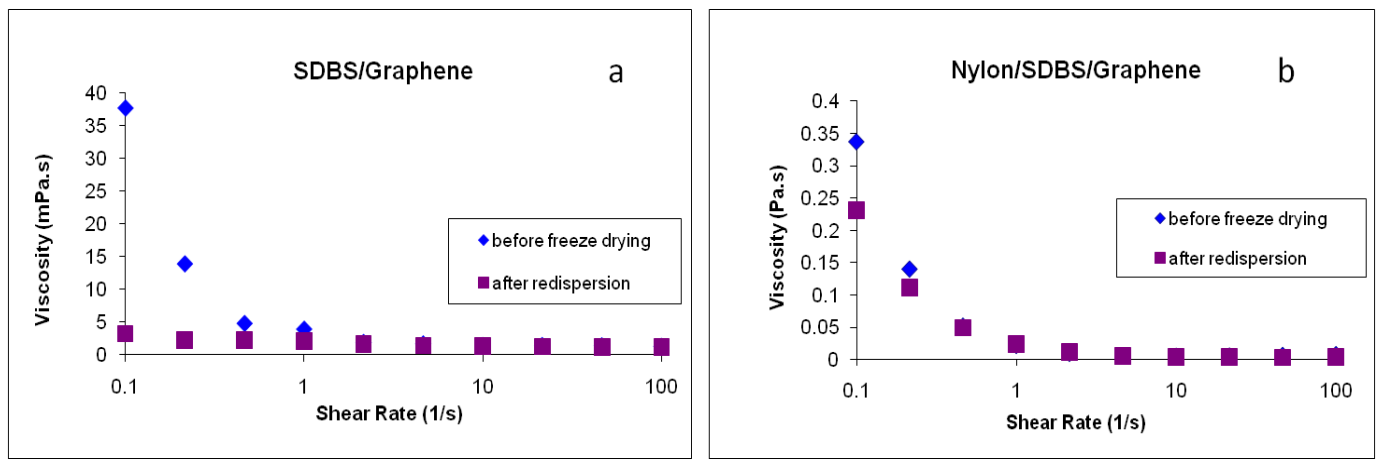

Fig 8: Rheological behavior of SDBS/graphene (a) before freeze drying and after redispersion, nylon/SDBS/graphene (b) before freeze drying and after redispersion. The viscosity of the polymer stabilized graphene dispersion shows a better recovery of the viscosity as opposed to the surfactant-stabilized graphene dispersion. The structure of the coated graphene is proved to be preserved during freeze-drying, as the data indicates no structural change.

# Localized in-situ polymerization on graphene surfaces for stabilized graphene dispersions

Sriya Das[*], Ahmed S. Wajid[*], John L. Shelburne[*], Yen-Chi Liao[*], Micah J. Green[*%]

[*]Department of Chemical Engineering, Texas Tech University, Lubbock, Texas 79409, USA

[%]Corresponding author: micah.green@ttu.edu

**Supplementary Information:**

**Absorbance spectra:**

We measure the absorbance spectra of the dispersions of surfactant-stabilized graphene, solvent-swelled surfactant stabilized graphene and surfactant-stabilized nylon-coated-graphene. Fig S1 shows the spectra for the three different samples. Fig. S2 shows the optical absorbance as a function of different concentrations of the graphene dispersions. We determined the extinction coefficient ($\alpha$) at a wavelength of 660nm using the linear relationship between the absorbance and calculated concentration for a particular dispersion ($A = \alpha l C$; where $l$ is the cell length). The extinction coefficient for the SDBS/graphene system is found to be 1660 mL mg$^{-1}$ m$^{-1}$.

**Importance of carbon tetrachloride:**

Note that the polymerization reaction does not require carbon tetrachloride. However, to aid the transportation of sebacoyl chloride to the hydrophobic interior of the surfactant micelles we add the organic solvent. To check whether carbon tetrachloride penetrates into the surfactant core we performed the following experiment. After heating the dispersion to 78° C (BP of $CCl_4$ is 76.7°C), the graphene aggregates. We believe that the heat destabilizes the surfactant micelles which lead to the aggregation of graphene. However, sonication for a minute after cooling stabilizes the graphene dispersion again.

**Ethyl acetate treatment:**

In the paper, the surfactant-stabilized nylon-coated graphene dispersions are treated with ethyl acetate to remove the surfactant from the system. The graphene dispersion is added to the ethyl acetate. After the treatment, the nylon-wrapped graphene dispersion destabilizes. This can be attributed to the fact that nylon itself is hydrophobic and cannot keep the graphene dispersed in water. SEM observations on the vacuum- filtered films and EA-treated vacuum -filtered films of surfactant stabilized graphene and nylon coated graphene are shown in Fig S3. A morphological difference is noted in case of Fig S3 (a) & (b). The EA treatment produces larger cracks in case of the only surfactant-stabilized graphene sample Fig S3 (c).In case of the EA treated nylon-coated graphene films, it can be predicted from Fig S3 (d) that the polymer is still present on the film due to the morphological difference of the surface.

**AFM Measurements:**

For the completeness of the results, more images of the surfactant-stabilized nylon-coated graphene are shown here (Fig S4).

**HRTEM Images:**

High-resolution TEM imaging shows that in both SDBS/graphene and SDBS/nylon/graphene the graphene is well dispersed and indeed single-to-few layers thick. (Fig S5). The nylon-coating and surfactant micelles are difficult to distinguish on the graphene surface, similar to Loyta *et al.* and Bourlinos *et al.*[1,2]. However, Fig. S5 (b) and Fig S5 (c) does show certain areas which may correspond to localized nylon coating.

**Rheology:**

Shear rheological experiments are performed to predict the dispersion quality from the relationship between the steady shear viscosity ($\eta$) and the shear rate ($\dot{\gamma}$). Fig. S6 (a) shows the response of the surfactant-stabilized dispersions before freeze-drying and after redispersion; there is little recovery of the original shear-thinning curve, which corresponds to a decrease in dispersion quality caused by the freeze drying. In Fig. S6 (b), the viscosity of the surfactant-stabilized nylon-coated graphene is depicted. The viscosity of the redispersed nylon-coated graphene shows only a small deviation from the initial values before freeze drying, in contrast to the surfactant-stabilized graphene. This implies that the dispersion quality in case of the surfactant-stabilized nylon-coated graphene after redispersion is fully recovered, with little structural change. We recommend the use of bulk rheological measurements as a characterization for dispersion stability; these measurements are analogous to prior rheological measurement to characterize dispersion stability and fluorescence measurements of SWNT dispersion quality before and after freeze-drying [3-6]. These rheological measurements also corroborate our AFM claims in Fig. 5c.

**Nylon-Graphene Composites:**

We use the freeze-dried EA-treated nylon-wrapped graphene as our filler in the nylon composites. We redisperse the freeze-dried powder along with nylon pellets (Sigma Aldrich, #429171) in 1,1,1,3,3,3-Hexafluoro-2-propanol (HFIP) (Sigma Aldrich, #52517). No visible aggregation occurs. We further evaporate the solvent to obtain the nylon-graphene composites (Fig S7 (a)). The composite is hot pressed to produce films (Fig S7 (b)). We are currently in the process of characterizing these composites using melt rheology; these developments will be reported in a rheology-focused future study (Fig S7 (c)).

**Supplementary figures:**

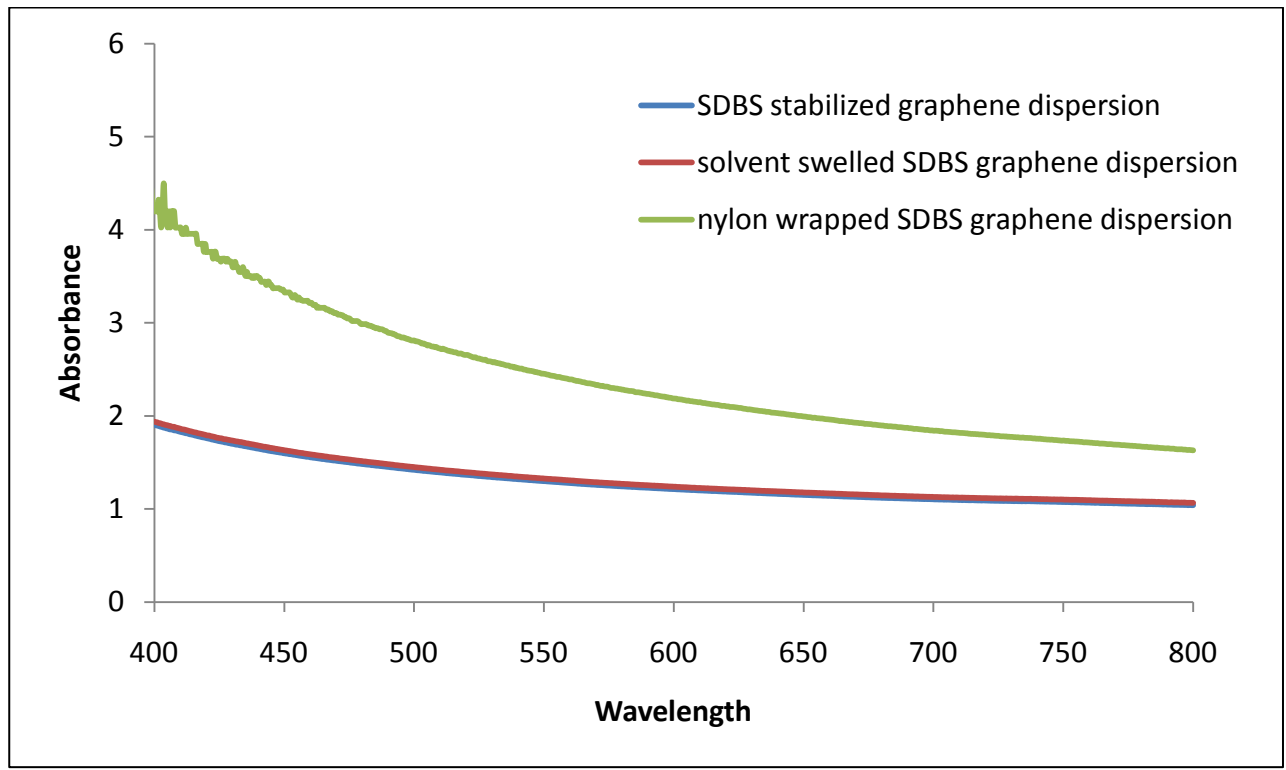

Figure S1: Background corrected absorbance spectra of (a) SDBS/graphene, (b) Solvent-swelled SDBS-coated graphene, (c) nylon/SDBS/graphene. The concentration of the centrifuged phase of the dispersion is determined by measuring the absorbance spectra.

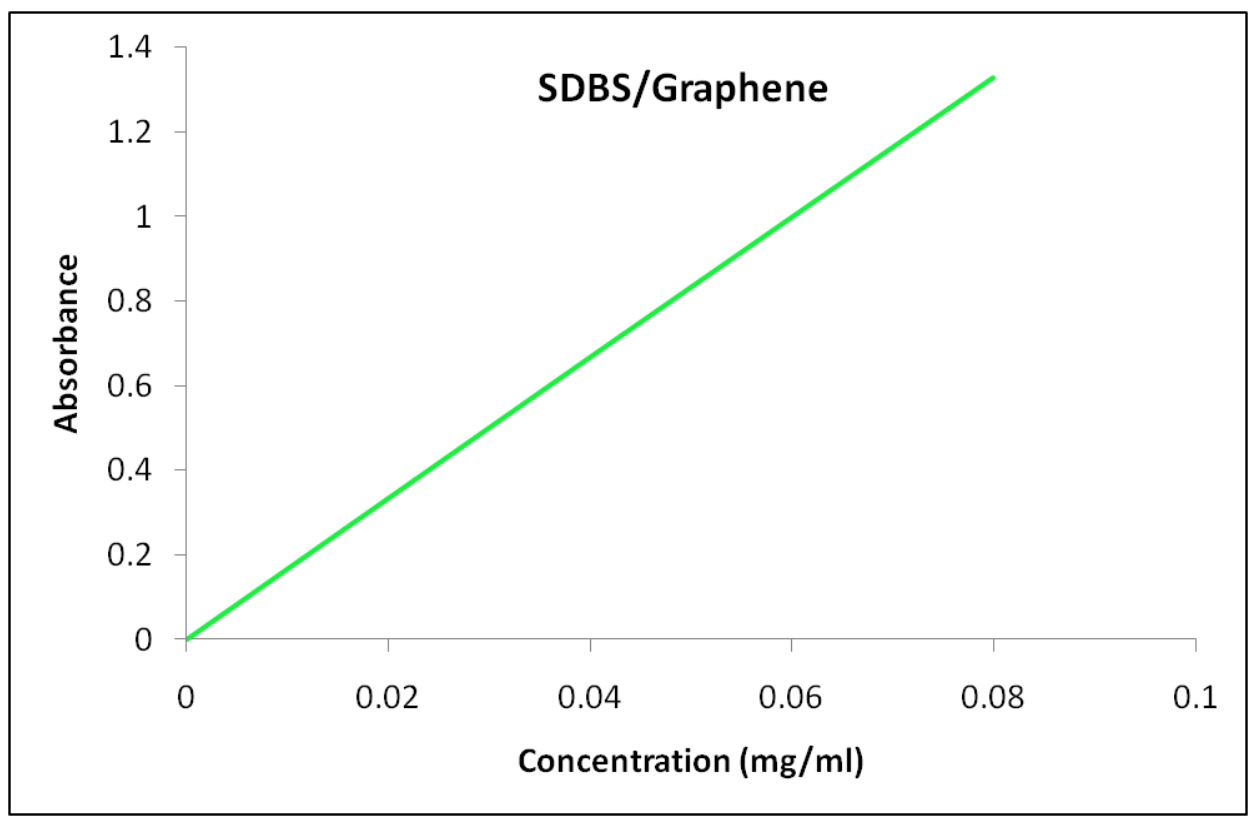

Figure S2: Optical absorbance as a function of different concentrations of SDBS/Graphene dispersion. The dispersion follows Lambert-Beer's law with an extinction coefficient of 1660 mL mg$^{-1}$ m$^{-1}$ at 660nm. The linear evolution of the data was utilized to obtain the concentration of graphene dispersion.

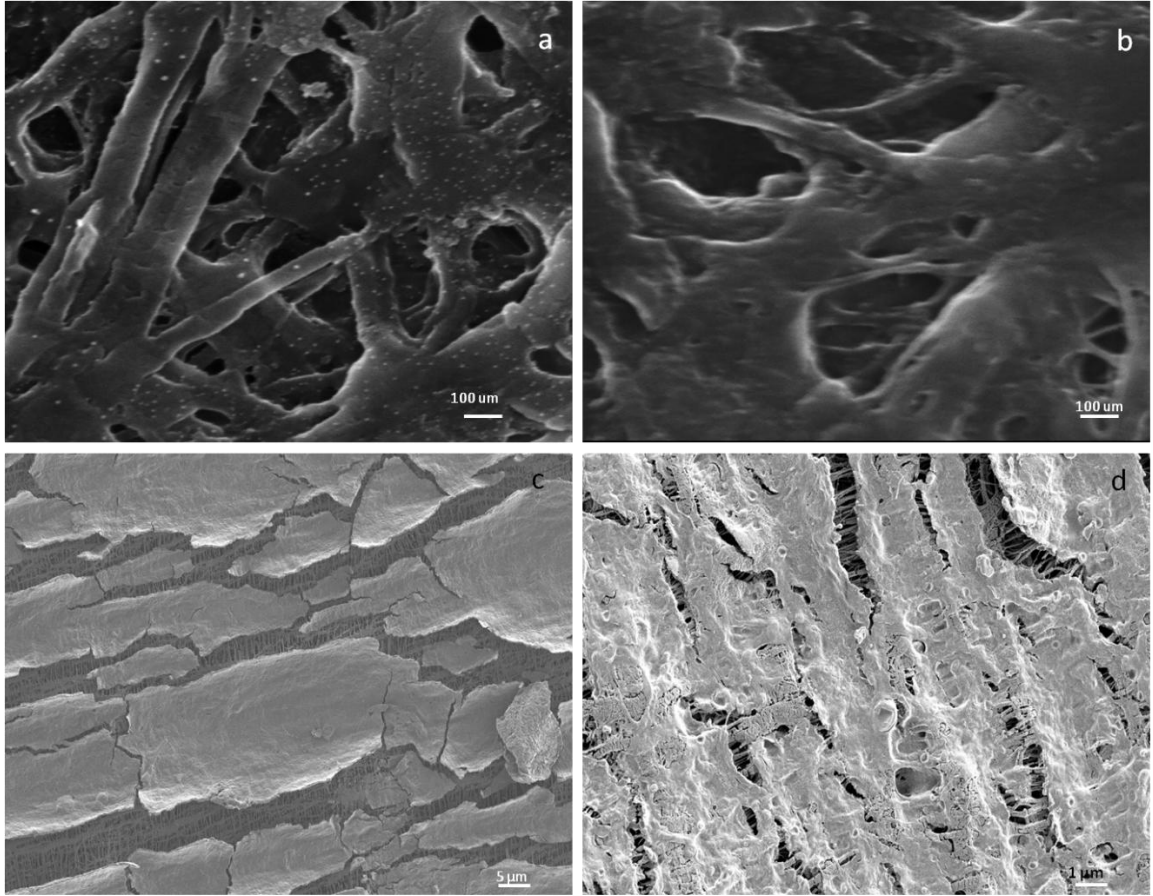

Figure S3: the SEM images of the vacuum filtered films of the (a) SDBS/graphene; (b) Nylon/SDBS/Graphene; (c) ethyl acetate treated SDBS/Graphene and (d) ethyl acetate treated Nylon/SDBS/Graphene is shown here. The EA treatment produces larger cracks in case of the surfactant-stabilized graphene sample. The morphological difference in case of the first two figures shows the presence of the polymer.

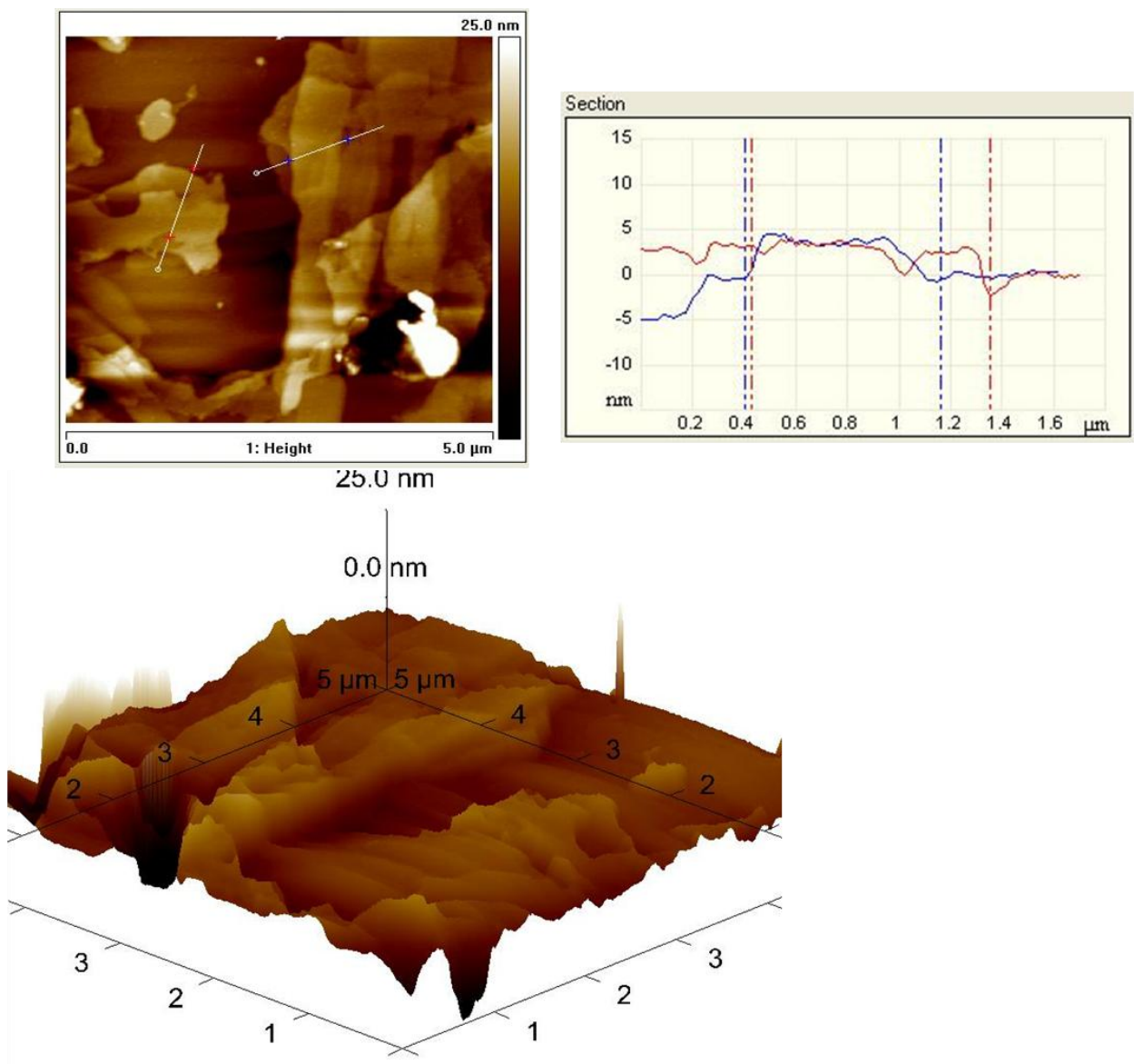

Figure S4: Additional images on the nylon-coated surfactant-stabilized graphene dispersion are shown here for completeness.

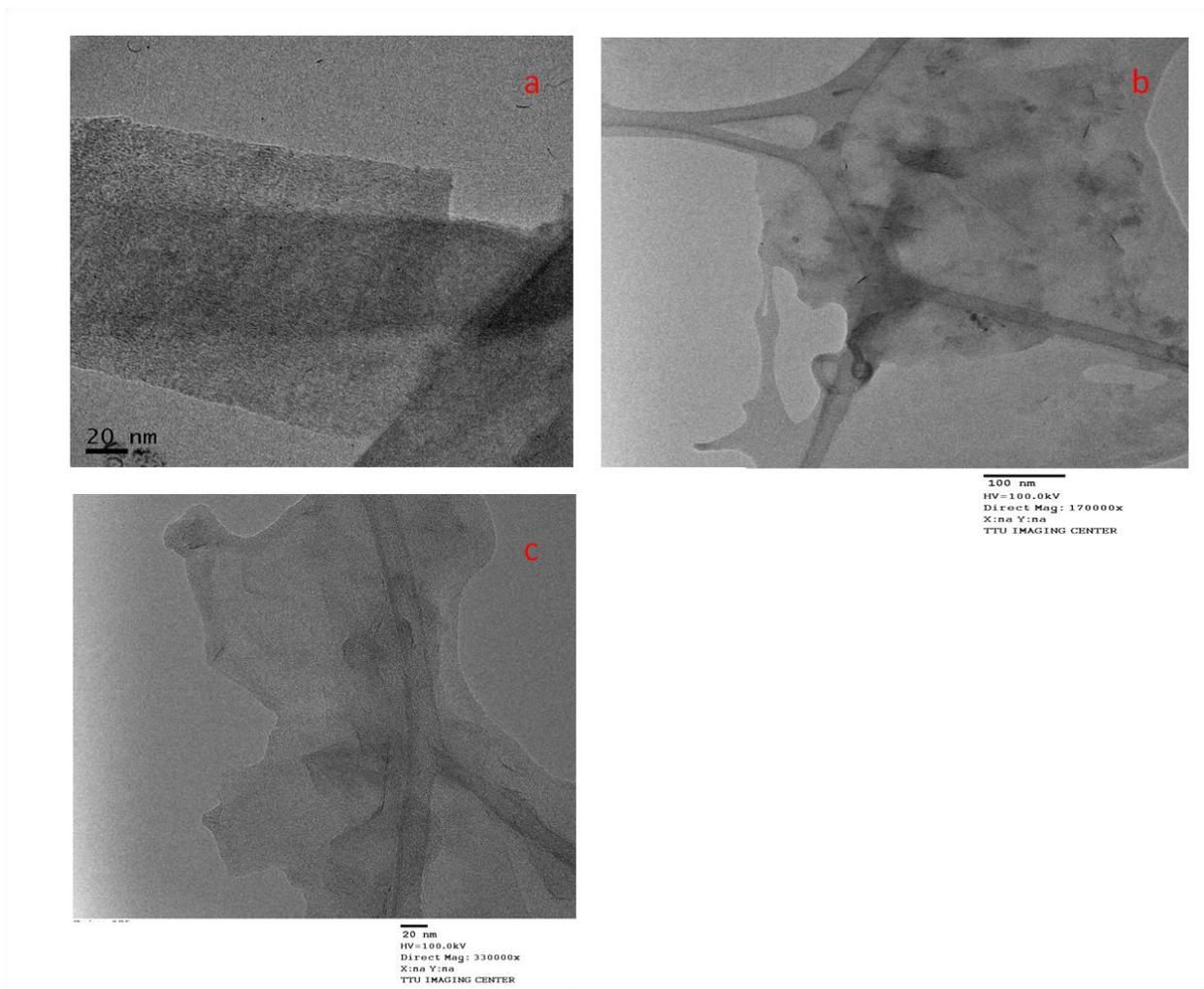

Figure S5: the TEM images of the (a) SDBS/graphene; (b) & (c) Nylon/SDBS/Graphene; the nylon-coating and surfactant micelles are difficult to distinguish on the graphene surface however some areas correspond to localized coating.

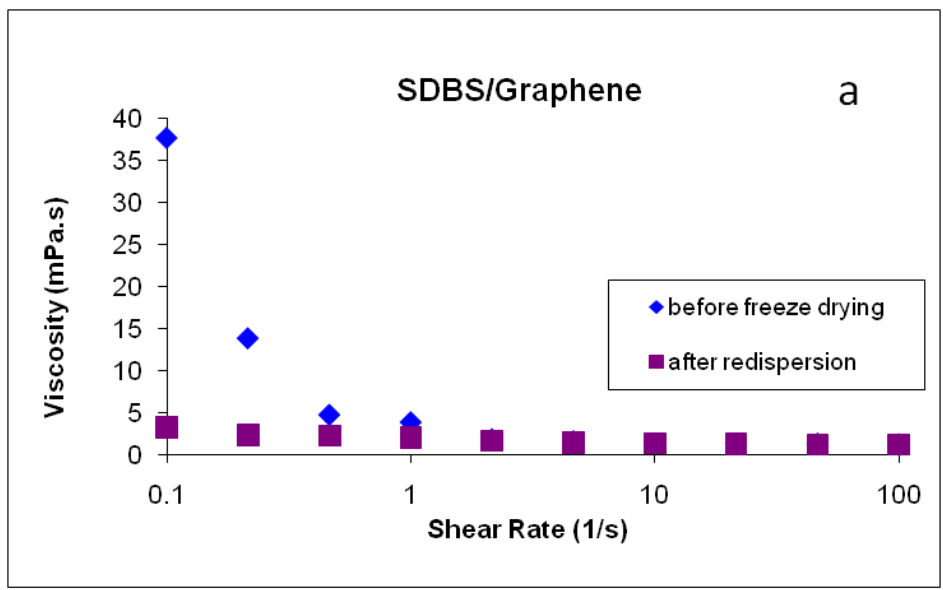

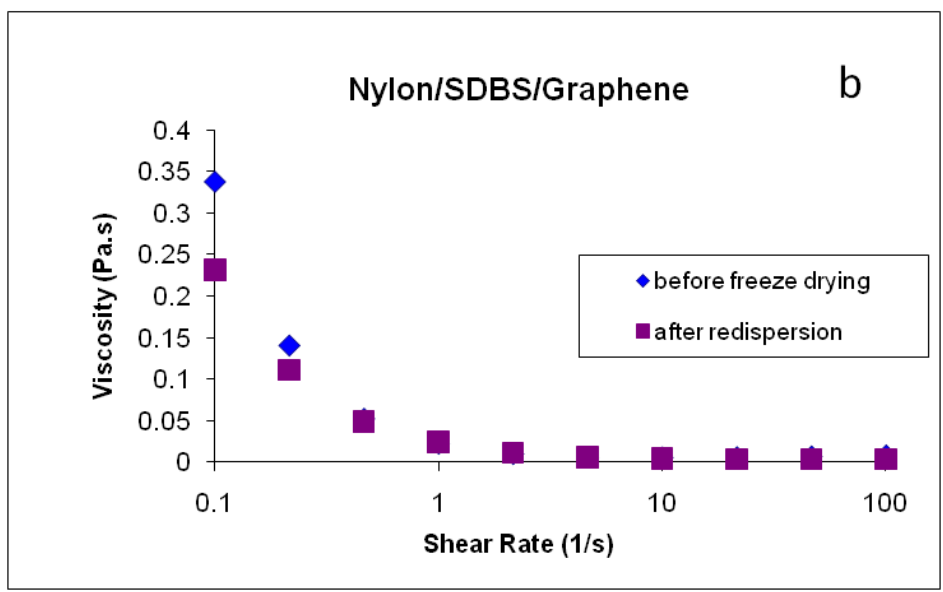

Fig S6: Rheological behavior of SDBS/graphene (a) before freeze drying and after redispersion, nylon/SDBS/graphene (b) before freeze drying and after redispersion. The viscosity of the polymer stabilized graphene dispersion shows a better recovery of the viscosity as opposed to the surfactant-stabilized graphene dispersion. The structure of the coated graphene is proved to be preserved during freeze-drying, as the data indicates no structural change.

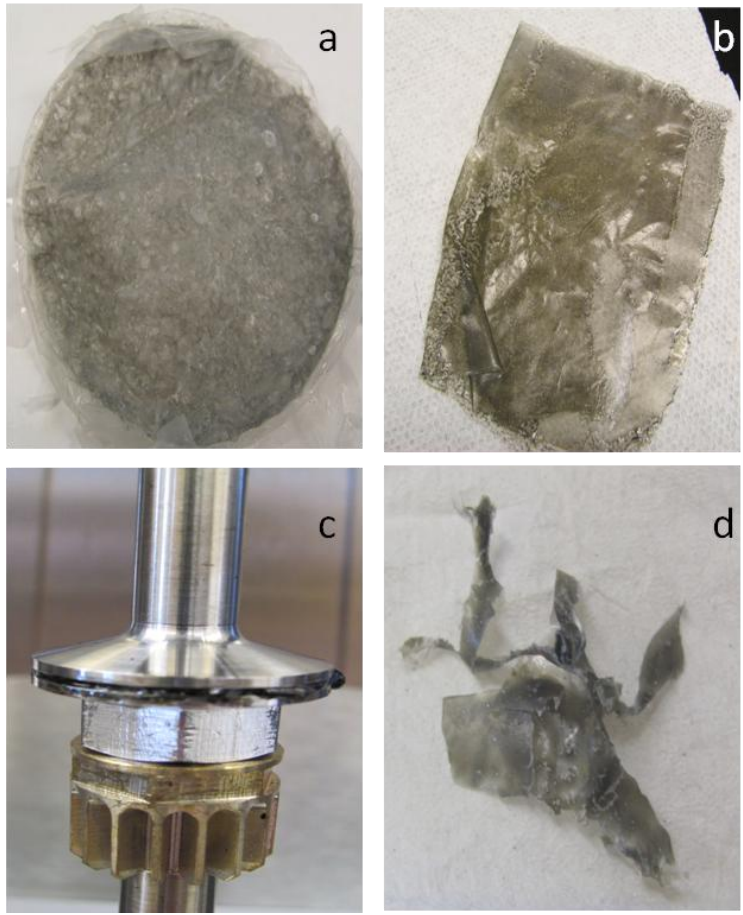

Figure S7: Digital camera images of (a) nylon composite loaded with nylon-coated graphene after solvent evaporation (graphene loading 0.5 wt%) (b) hot pressed film of nylon composite loaded with nylon-coated graphene (c) melt rheology sample of nylon composite loaded with nylon-coated graphene (d) post-melt nylon composite loaded with nylon-coated graphene.